\documentclass[]{interact}

\usepackage{epstopdf}
\usepackage{subfigure}

\usepackage{natbib}
\usepackage{lineno}
\usepackage{color}
\usepackage{ulem}
\usepackage{soul,xcolor}
\setstcolor{red}

\bibpunct[, ]{(}{)}{;}{a}{}{,}
\renewcommand\bibfont{\fontsize{10}{12}\selectfont}

\theoremstyle{plain}

\theoremstyle{definition}

\theoremstyle{remark}

\begin{document}

\articletype{RESEARCH ARTICLE}

\title{An affordable and customizable wave buoy for the study of wave-ice interactions: design concept and results from field deployments}

\author{
\name{
Tsubasa Kodaira \textsuperscript{a}\thanks{CONTACT Tsubasa Kodaira. Email: kodaira@edu.k.u-tokyo.ac.jp}, Tomotaka Katsuno\textsuperscript{a}, Takehiko Nose\textsuperscript{a}, Motoyo Itoh\textsuperscript{b}, Jean Rabault\textsuperscript{c}, Mario Hoppmann\textsuperscript{d}, Masafumi Kimizuka\textsuperscript{e}, and Takuji Waseda\textsuperscript{a}}
\affil{
\textsuperscript{a}Department of Ocean Technology, Policy, and Environment, Graduate School of Frontier Sciences, The University of Tokyo, Kashiwa, Japan \\
\textsuperscript{b}Institute of Arctic Climate and Environment Research, Japan Agency for Marine-Earth Science and Technology, Yokosuka, Japan \\ 
\textsuperscript{c}{IT Department, Norwegian Meteorological Institute, Oslo,Norway} \\
\textsuperscript{d}{Alfred-Wegener-Institut, Helmholtz-Zentrum für Polar- und Meeresforschung, Bremerhaven, Germany} \\
\textsuperscript{e}{Department of Mechanical Systems Engineering, Tokyo Metropolitan College of Industrial Technology, Tokyo, Japan}}}

\maketitle


\begin{abstract}
In the polar regions, the interaction between waves and ice has a crucial impact on the seasonal change in the sea ice extent. However, our comprehension of this phenomenon is restricted by a lack of observations, which, in turn, results in the exclusion of associated processes from numerical models. In recent years, availability of the low-cost and accurate Inertial Motion Units has enabled the development of affordable wave research devices. Despite advancements in designing innovative open-source instruments optimized for deployment on ice floes, their customizability and survivability remain limited, especially in open waters. This study presents a novel design concept for an affordable and customizable wave buoy, aimed for wave measurements in marginal ice zones. The central focus of this wave buoy design is the application of 3D printing as rapid prototyping technology. By utilizing the high customizability offered by 3D printing, the previously developed solar-powered wave buoy was customized to install a battery pack to continue the measurements in the high latitudes for more than several months. Preliminary results from field deployments in the Pacific and Arctic Oceans demonstrate that the performance of the instruments is promising. The accuracy of frequency wave spectra measurements is found to be comparable to that of considerably more expensive instruments. Finally, the study concludes with a general evaluation of using rapid prototyping technologies for buoy designs and proposes recommendations for future designs.
\end{abstract}


\section{Introduction}
Direct observations of wave-ice interactions are necessary due to their potential significance for improving future projections of the Arctic and Earth's climate. Wind waves have the potential to decrease the sea ice extent via dynamic and thermodynamic processes. As incoming waves break up the sea ice, the export of sea ice to adjacent, warmer areas may be enhanced. Previous observational studies have shown that storm-generated ocean waves can propagate and break sea ice for hundreds of kilometers \citep{kohout2014,Collins2015}. 

The breaking up of sea ice can cause the floe size to reduce, leading to increased heat transfer with adjacent seawater \citep{Steele1992, Toyota2011}. Moreover, over the Arctic Ocean with a subsurface temperature maximum, energetic turbulence in the upper ocean can cause sea ice melt. A significant retreat of the sea ice-covered area was thought to be caused by vertical mixing in the upper ocean under stormy conditions \citep{Smith2018}. The effect of surface waves on sea ice formation has been less explored. 

Previously, \cite{Kohout2015} designed an accelerometer-based device for measuring the wave-induced motion of ice floes to understand the complex physical processes of wave-ice interactions. More recently, \cite{Rabault2022} presented an open-source drifter and wave-monitoring instrument using a low-cost GNSS receiver and IMU sensor. This device could provide an order of magnitude more observational data under a specific instrumentation budget. Deploying multiple wave sensors can be an effective observation method for studying wave-ice interactions due to the significant spatial variation of sea ice and wave conditions in the marginal ice zone. To further study this phenomenon, a floating platform to enclose the device is essential and effective to extend the area of measurements to open water as demonstrated by \cite{Nose2023}.

Recently, there have been significant technical advancements in drifting-type wave measurement buoys. Notably, the compact and solar-powered Spotter directional wave buoy, developed by Sofar Technologies, Inc. \citep{Raghukumar2019}, stands out as a prominent example, with the established global wave measurement network. Another noteworthy innovation is the microSWIFT, named after its predecessor, the Surface Wave Instrument Float with Tracking (SWIFT) buoy \citep{Thomson2012}. Despite its small size, close to a pet bottle, the microSWIFT weighs merely 0.9 kg and successfully measures directional wave fields \citep{Rainville2023}.

Given that these wave buoys provide directional wave spectra, the data obtained are invaluable for conducting scientific research and enhancing numerical wave models. Other academic groups have also contributed to the development of original wave buoys such as those by \cite{Hirakawa2016, Yaozhao2022}. Furthermore, it is noteworthy that some of these drifting wave buoys were used for the wave-ice interaction studies. Spotter wave buoys were used to sutdy the wave attenuation under the grease ice \citep{Kodaira2021}. 
SWIFT buoys have been effectively employed in the marginal ice zone in the Arctic to study wave groups in Pancake sea ice \citep{Thomson2019}.

We have developed an affordable, compact (20~cm), solar-powered wave buoy named FZ \citep{Kodaira2022}. A distinctive feature of FZ, distinguishing it from the previously mentioned wave buoys, is its hull. The hull is fabricated using 3D printing technology, which significantly reduces the cost of the buoy hull to as low as 60 USD. Moreover, being categorized under Rapid Prototyping technology, 3D printing grants FZ exceptional customizability.

This paper showcases the high level of customizability by presenting the redesign of FZ wave buoy for deployments in the marginal ice zone, leading to the development of eXpendable FZ (XFZ). Furthermore, we assess the performance of the deployed wave buoys based on the results of two dedicated field experiments, where a total of 14 XFZ instruments were deployed. We also address the identified limitations and offer recommendations for future design enhancements.

\section{Buoy Design}
\subsection{Concept}
Our fundamental principle in designing the wave buoy was to develop a cost-effective and customizable device in both hardware and firmware. This concept is crucial for accelerating the trial-and-error process and promoting rapid development. This approach is particularly suitable for small research groups studying scientific topics in the early stages where established methods have not yet been developed. To achieve enhanced affordability and customizability, we employed the Rapid Prototyping technology of a 3D printer. The utilization of a 3D printer for fabricating the wave buoy hull ensures minimal production costs, while maximizing the potential for customization, as described  below.

\begin{figure}[h]
\centering
\resizebox*{14cm}{!}{\includegraphics{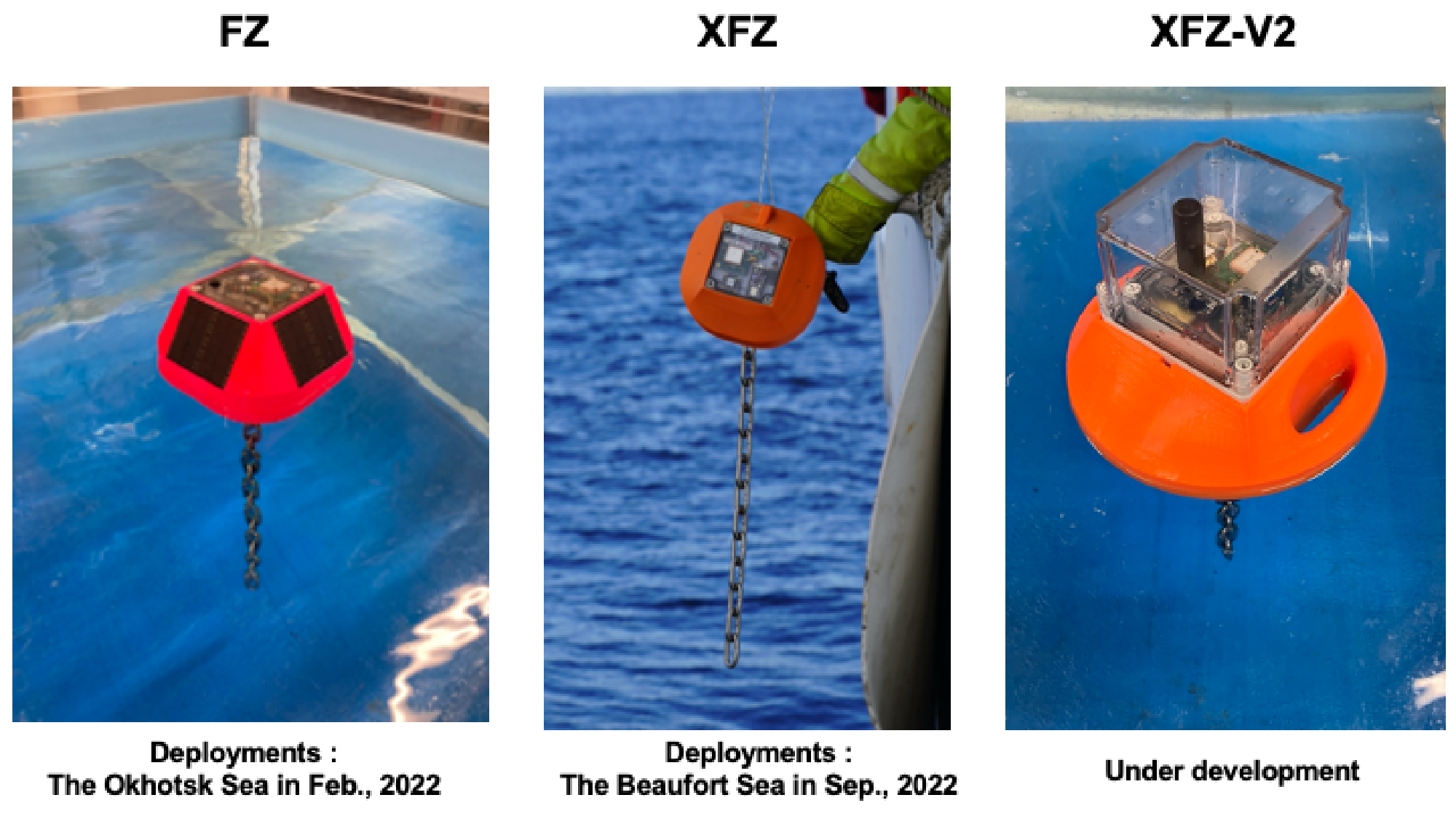}}
\caption{Customizing process of the wave buoy FZ} 
\label{figure0}
\end{figure}

Based on the conceptual framework, we have undertaken the customization of our initial wave buoy, FZ (Figure~\ref{figure0}), for its deployment in high latitudes. The primary modification is the  transition in the power supply from solar power to Lithium primary batteries, as solar radiation is limited in these regions. In order to install the relatively substantial battery volume, an additional water-tight box was integrated to supply power to the sensor box. For this enhancement of the longer battery life, the buoy hull underwent a redesign, resulting in a slightly larger volume to accommodate extra mounting space and buoyancy. The redesigned wave buoy, depicted in Figure~\ref{figure0}, has been named as eXpendable FZ (XFZ). Notably, the employment of a 3D printer accelerated the redesign process.

Building upon XFZ, XFZ-V2 has undergone further customization, guided by the results of the field trials presented in this paper. The sensor box was encapsulated in an additional, slightly larger, watertight enclosure, to create a robust double-hull structure ( Figure~\ref{figure0}). This customization aims to enhance the wave buoy's physical resilience against potential damages caused by seawater intrusion and collisions with sea ice. Additionally, an external antenna was incorporated to improve the Iridium satellite communication capabilities. These illustrative examples of the customization affirm the adaptable nature and the approach is promising for investigating wave-ice interactions.

As partly described above, we utilized a 3D printer for the fabrication of the wave buoy. However, a notable challenge with 3D printed components used in ocean-sensing devices was their lack of watertight integrity. To overcome this issue and ensure the buoyancy and protection of the electronics, we employed commercially available watertight boxes with an International Protection (IP) rating of IP66/67 in conjunction with the 3D printer. The hull of the buoy is thoughtfully designed to securely enclose the aforementioned boxes. The shell-like structure of the hull is designed in order that polyurethane form can be injected to  fill in the void of the hull for secure buoyancy (see, Appendix for further information).  This way of integration of a watertight box with the 3D printing technology is a novel approach, which allowed us to utilize the rapid prototyping and produce a water-resistant product while preserving a high degree of adaptability.
 
Selection of the microcontroller and programming language is also made from a rapid prototyping perspective. The Adafruit Feather nRF52840 Sense was chosen as the microcontroller because (1) some environmental sensors are embedded on the board and (2) a high functional scalability is supported by a series of extension boards. For example, as demonstrated below a data logging function can be added with SD-card data logging extension board when the device can be retrieved. As the programming language for XFZ, CircuitPython, a derivative of MicroPython, was chosen  because it does not require a compilation process and enables fast programming, thanks to the embedded programming language with disk space. Compared to C language, CircuitPython allows for shorter code. For XFZ, the total number of lines in the codes we created was less than 400.

Other design concepts, such as scalability and sustainability, were considered less important. However, scalability becomes necessary in cases where there is a massive deployment of sensors, as seen in recent studies that deployed several hundred ocean surface drifters \cite{Poje2014, Dasaro2018}. In the near future, sustainability must also be incorporated into the design philosophy. Since most Lagrangian-type drifting ocean sensors are expendable, materials with lower environmental impacts, such as biodegradable plastic, should be chosen, following the previous study by \cite{Novelli2017}.

\subsection{Wave measurement and on-board analysis}
The measurement cycle is programmed as follows: first, the GNSS signal is acquired, which usually takes a minute to obtain location data. Next, IMU data is sampled for 1024 s to measure waves. The spectral analysis, described in detail in the next paragraph, is conducted on the vertical acceleration after applying the tilt correction. Finally, the data is transferred via Iridium satellite. The measurement interval is usually set hourly but can be changed using Iridium satellite communication after deployment.

The analysis program implemented on board for frequency wave spectra of ocean surface waves is similar to the previous studies \citep{Kohout2015, Rabault2022}. Accelerations and absolute orientations are measured for a duration of 1024 seconds at a sampling rate of 64 Hz to obtain wave measurements. Following this, the data is downsampled to 4Hz after averaging over a period of 0.25 seconds. The time series of the vertical acceleration was subsequently calculated following \cite{Bender2008} for estimating the power spectral density. The time series was divided into four segments, each having a length of 256 seconds. For every segment, the Hanning window was applied before performing the fast Fourier transform, with the aim of reducing spectral leakage. The periodogram estimates were obtained and then averaged. To minimize the volume of satellite data transfer, the spectral estimates were interpolated to an irregular interval defined by a frequency array $f_i [i]=(1.06)^i/36,   (0\leq i \leq 61)$. The interpolation was performed after applying the three-point moving average on the estimated power spectral density.

Finally, the power spectral density, which has been interpolated, is transmitted through Iridium SBD. The transmitted data also comprises information about the measurement location and conditions, such as battery voltage, board temperature, air humidity, and air pressure. In case the data transmission fails after certain retries, the data is temporarily stored in RAM, and the transmission is retried during the next measurement cycle.

\subsection{Postprocessing}
The power spectral density of surface waves is calculated at the postprocessing stage by multiplying $(2\pi f)^{-4}$ to the power spectral of the vertical acceleration transmitted via Iridium SBD. The multiplication of $(2\pi f)^{-4}$ means double integration in time in the frequency space. In the low-frequency part, the calculated wave displacement spectral density often follows the form of $\epsilon(2\pi f)^{-4}$ where $\epsilon$ is the constant. These are not likely the wave signal but the manifestation of the white noise in the acceleration measurements.

Following the previous studies \citep{Waseda2018,Nose2023}, an ideal high-pass filter is thus applied by specifying the cut-off frequency for each result of the power spectral. The cut-off frequency is set to the frequency where the power spectral density takes the local minimum. The specification of the cut-off frequency is done after the high pass filtering by taking the moving average of the obtained power spectral density. 

To calculate the bulk wave parameters, the n\textsuperscript{th} spectral moment is defined as 
\begin{equation}
 m_n=\int_{f_1}^{f_2} f^n S(f) df
\end{equation}
where $S(f)$ is the estimated PSD of the surface wave, $f_1$ is the cut-off frequency determined by the spectral shape, and $f_2$ is constant. The significant wave height $H_s$, and the mean wave period $T_{m01}$ is then defined as follows, respectively, 
\begin{equation}
 H_s=4 \sqrt{m_0}
\end{equation}
\begin{equation}
 T_{m01}= m_o/m_1
\end{equation}
As introduced below, the measurements of XFZ are compared to Spotter wave buoys for validation. For the comparison, the same frequency range from $f_1$ to $f_2$ is used for both the XFZ and Spotter wave buoy to compare these bulk wave parameters.

\section{Results from Field Trials}
To obtain a comprehensive understanding of wave-ice interaction, measurements in both open water and sea ice-covered regions are necessary. This section reports the results of three different field deployments of XFZ: (1) to measure typhoon conditions, (2) in open waters of the Beaufort Sea, and (3) on an ice floe in the Greenland Sea.

\subsection{Wave buoy under Typhoon in the Pacific Ocean}
On August 11th, 2022, Typhoon Meari emerged in the southern region of Japan and subsequently made landfall in Shizuoka prefecture on August 13th. Its projected trajectory closely intersected with the course of R/V Mirai that was headed towards the Arctic. Despite being a comparably weak typhoon, as evidenced by the minimum sea level pressure of merely 996 hPa, a relatively elevated sea state accompanied by a significant wave height of 5 m was anticipated. The typhoon event was deemed a propitious opportunity to assess the resilience of the devised wave buoy.

\begin{figure}
\centering
\subfigure{\resizebox*{8.5cm}{!}{\includegraphics{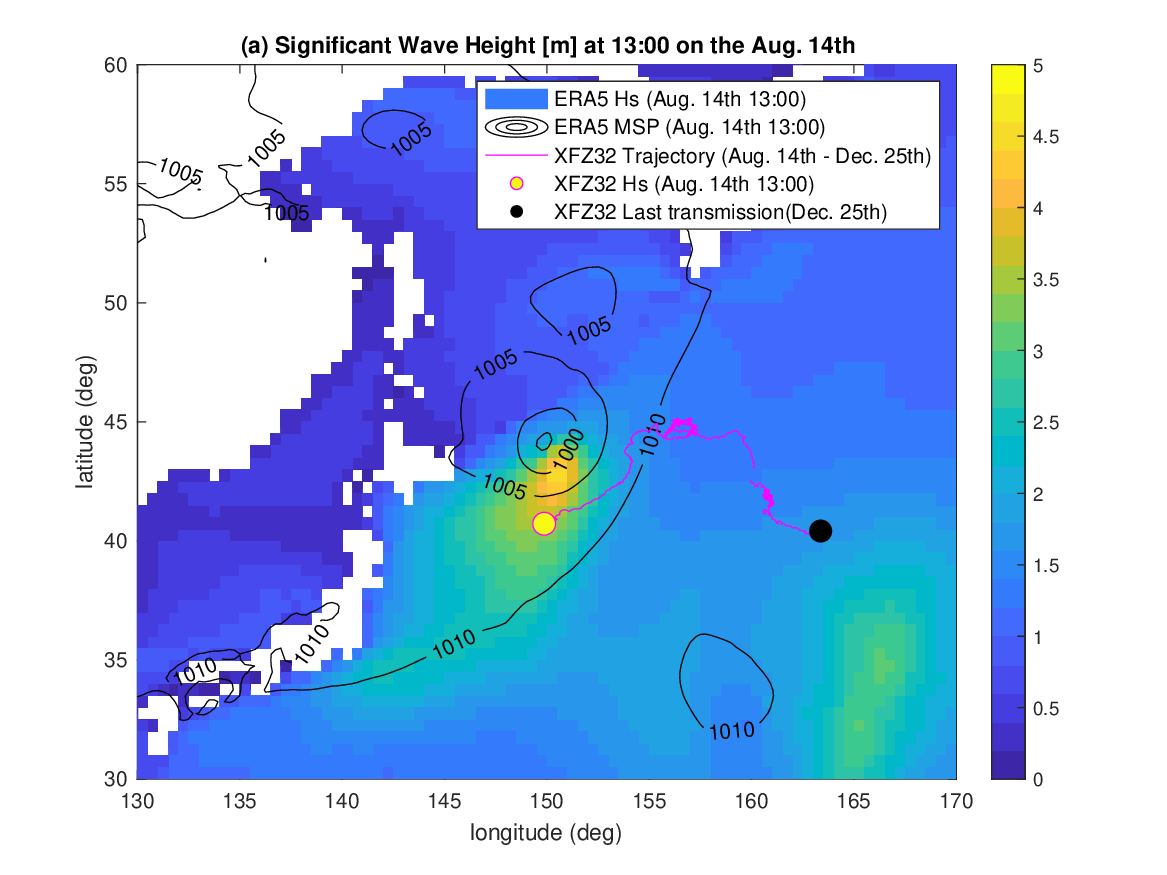}}}
\hspace{0pt}
\subfigure{\resizebox*{5.5cm}{!}{\includegraphics{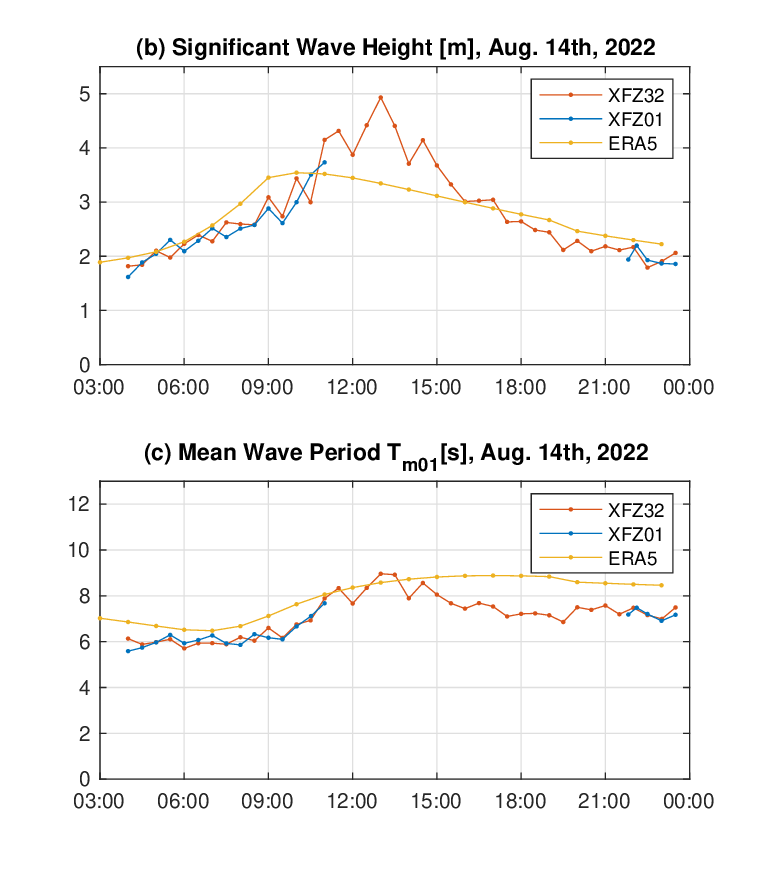}}}
\caption{
(a) Significant wave height when XFZ32 recorded the maximum significant wave height associated with typhoon Meari passing east of mainland Japan. The background color shows the estimates by ERA5. The black lines indicate the mean sea level pressure based on ERA5. The magenta line shows the trajectories of XFZ32 by the time of the last transmission on December 25th.(b) time series of the significant wave height, and (c) mean wave period $T_{m01}$.
} 
\label{figure4}
\end{figure}

In advance of the approach of Typhoon Meari, two XFZs, namely XFZ32 and XFZ01, were deployed from R/V Mirai in the northwest Pacific (36.22'N, 9°48.01'E) at 3:05 UTC on August 14th. The sampling interval was set to 30 minutes. The average wind speed during the deployment was approximately 14.5 m/s, at a height of 25 m above sea level. Following the deployment, XFZ32 observed a significant increase in the measured significant wave height, reaching a maximum of 5 m at 13:00 on August 13th (Figure~\ref{figure4}). Additionally, the mean wave period increased to reach the peak value of 9 seconds at 13:00 on August 13th. On the other hand, XFZ01 ceased transmission for unknown reasons after 11:00, only to resume at 22:00. Since XFZ01 restarted the transmission, the buoy may not have been significantly damaged physically. Since it stopped under the relatively high sea state conditions, some problems in assembly such as loose connection are possible.

A comparison between the calculated significant wave height and mean wave period with ERA5 \citep{Hersbach2020} reanalysis reveals that ERA5 failed to predict the magnitude and timing of the peak. It is noteworthy that ERA5 has a tendency to underestimate extreme wave conditions, as reported by \cite{Kodaira2023}. However, XFZ32 measured a significant wave height of approximately 5 m, for which condition ERA5 is supposed to perform well. 

\begin{figure}
\centering
\resizebox*{14cm}{!}{\includegraphics{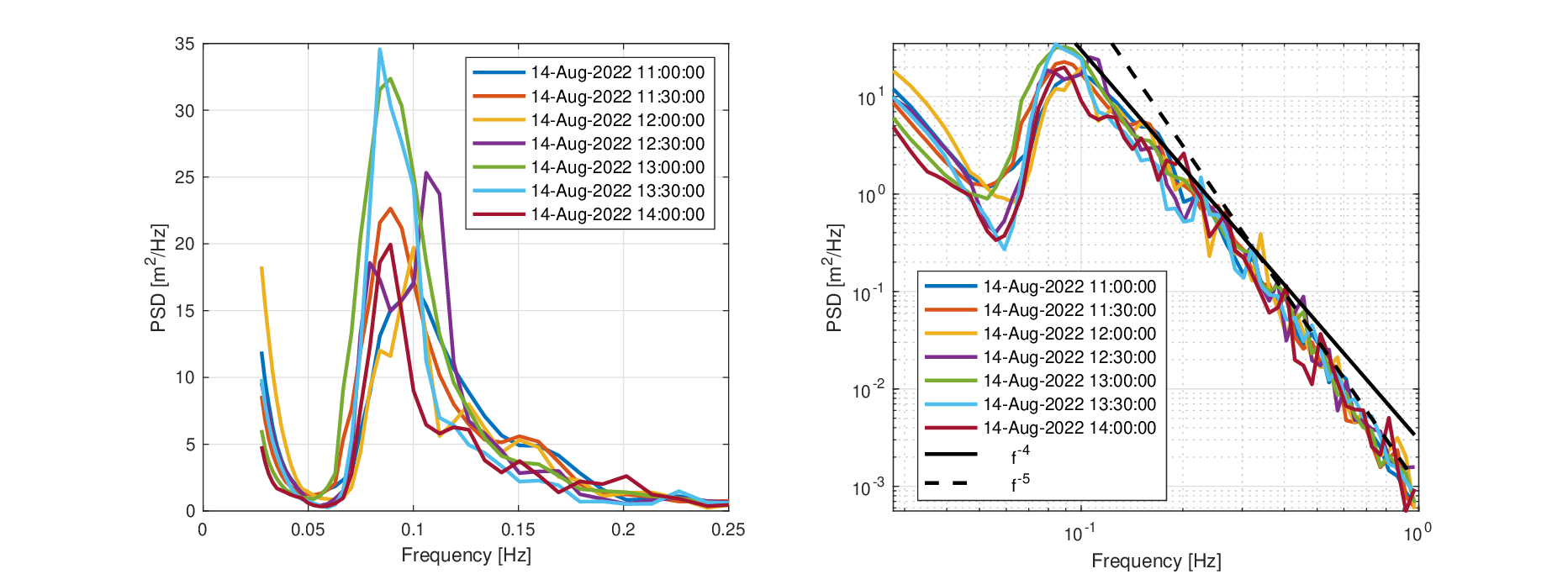}}
\caption{
Power spectral densities obtained from XFZ32 wave buoy while traversing Typhoon Meari.  The linear and logarithmic scale representations are depicted in the left and right panels, respectively.
} 
\label{figure5}
\end{figure}

XFZ32 also captured the evolution of frequency spectra during the traverse of Typhoon Meari. Figure~\ref{figure5} displays the acquired frequency spectra from 11:00 to 14:00.  In addition to the linear plot, a logarithmic plot of the power spectra was also generated to expand the range of observation.  It is evident that wind waves are detected up to 1 Hz, which is made possible by the compact size of the XFZ.

The spectral estimates from raw data appear to have a relatively substantial degree of uncertainty.
\begin{equation}
 \frac{S(f) \times DoF}{\chi^2 (DoF,\frac{1-\alpha}{2})} \leq 
 S(f)  \leq 
 \frac{S(f) \times DoF}{\chi^2 (DoF,\frac{1+\alpha}{2})}
\end{equation}
where $\chi^2$ are percentage points of a chi-square probability distribution and $\alpha$ defines the confidence interval \citep{Earle1996,Kohout2015}. Based on this DoF, the 90 $\%$ confidence interval can be computed between 0.64 $S(f)$ and 1.80 $S(f)$. It is thus not conclusive whether the temporal change in the frequency spectra indicates the rapid changes in waves under the typhoon passage or the uncertainty in the spectral estimates.

The data collected from XFZ32 and XFZ01 after the passage of typhoon Meari indicated that both wave buoys survived. Despite the weak intensity of the typhoon, the survival of both wave buoys was a positive sign for future studies on waves under typhoons. Before transmission was interrupted, XFZ32 made a total of 3975 measurements over a period of 165 days, while XFZ01 only made 215 measurements. 

\subsection{Wave buoy array in the Beaufort Sea}
\begin{figure}
\centering
\subfigure[]{\resizebox*{7cm}{!}{\includegraphics{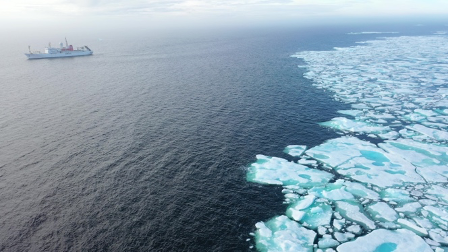}}}
\hspace{1pt}
\subfigure[]{\resizebox*{5cm}{!}{\includegraphics{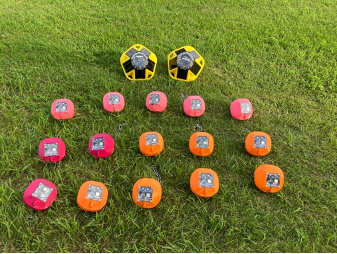}}}
\caption{ (a) Photograph taken by Tomotaka Katsuno showing the sea ice edge where wave buoys were deployed from R/V Mirai in MR22 cruise. (b) Wave buoys prepared for the MR22 cruise. Although only two Spotter buoys are shown in the photo, three Spotter buoys were deployed.
} 
\label{figure6}
\end{figure}

\begin{figure}
\centering
\resizebox*{14cm}{!}{\includegraphics{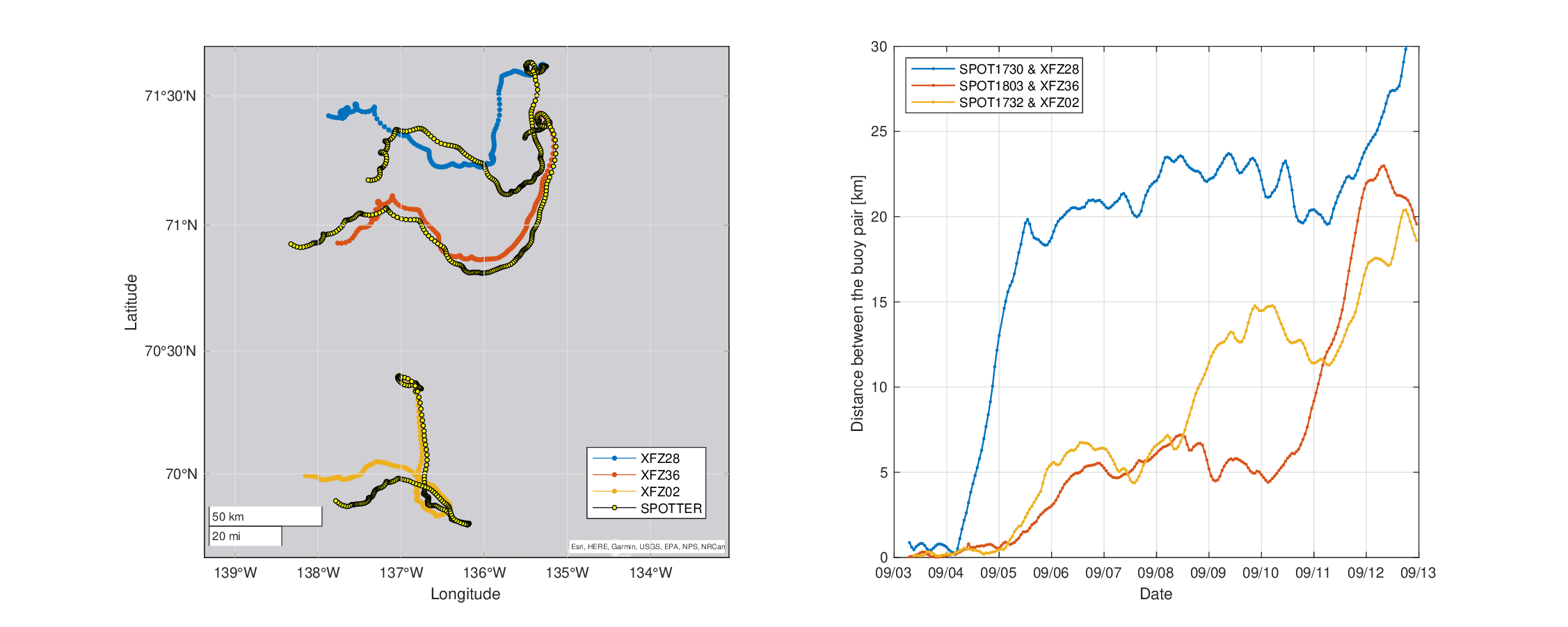}}
\caption{
Trajectories of the three pairs of XFZ and Spotter wave buoys for 10 days from September 3rd (left). Distance between the buoy pair (right).
} 
\label{figure7}
\end{figure}

During the MR22 expedition on board R/V Mirai, the vessel sailed towards the sea ice edge in the Beaufort Sea on September 3rd (Figure~\ref{figure6}). Upon departure from the sea ice edge, twelve XFZs were deployed along the ship's course with a spatial interval ranging from 10 km to 20 km. Among the twelve deployments, Spotter wave buoy was also deployed alongside XFZ at three different locations, as illustrated in Figure~\ref{figure7}. The wave measurements obtained by the two types of wave buoys were compared for a duration of ten days after the deployment. However, the comparison was limited to only ten days due to two reasons: (1) the distance between the XFZ and Spotter wave buoys increased with time and (2) the occurrence of a software clock issue\footnote{The software issue became influential on the results after about two weeks of measurement, in which the chosen software clock has only 22 bits of precision and loses precision with time. The problem was readily solved by choosing another software clock that uses arbitrarily long integers and the improvement was confirmed based on the results of the next deployments near the Antarctica in December 2022 (Figure~\ref{figure0}).}

The initial deployment of the Spotter and XFZ buoys took place at the sea ice edge, while the second and third pairs were situated at distances of 25 km and 150 km from it, respectively. After deployment, all the buoys drifted generally westward by the oceanic Beaufort Gyre, and their trajectories gradually diverged over time due to varying leeway characteristics and oceanic turbulence. Notably, the first pair deployed at the sea ice edge showed a significantly faster separation speed compared to the other pairs (Figure~\ref{figure7}).

\begin{figure}
\centering
\resizebox*{14cm}{!}{\includegraphics{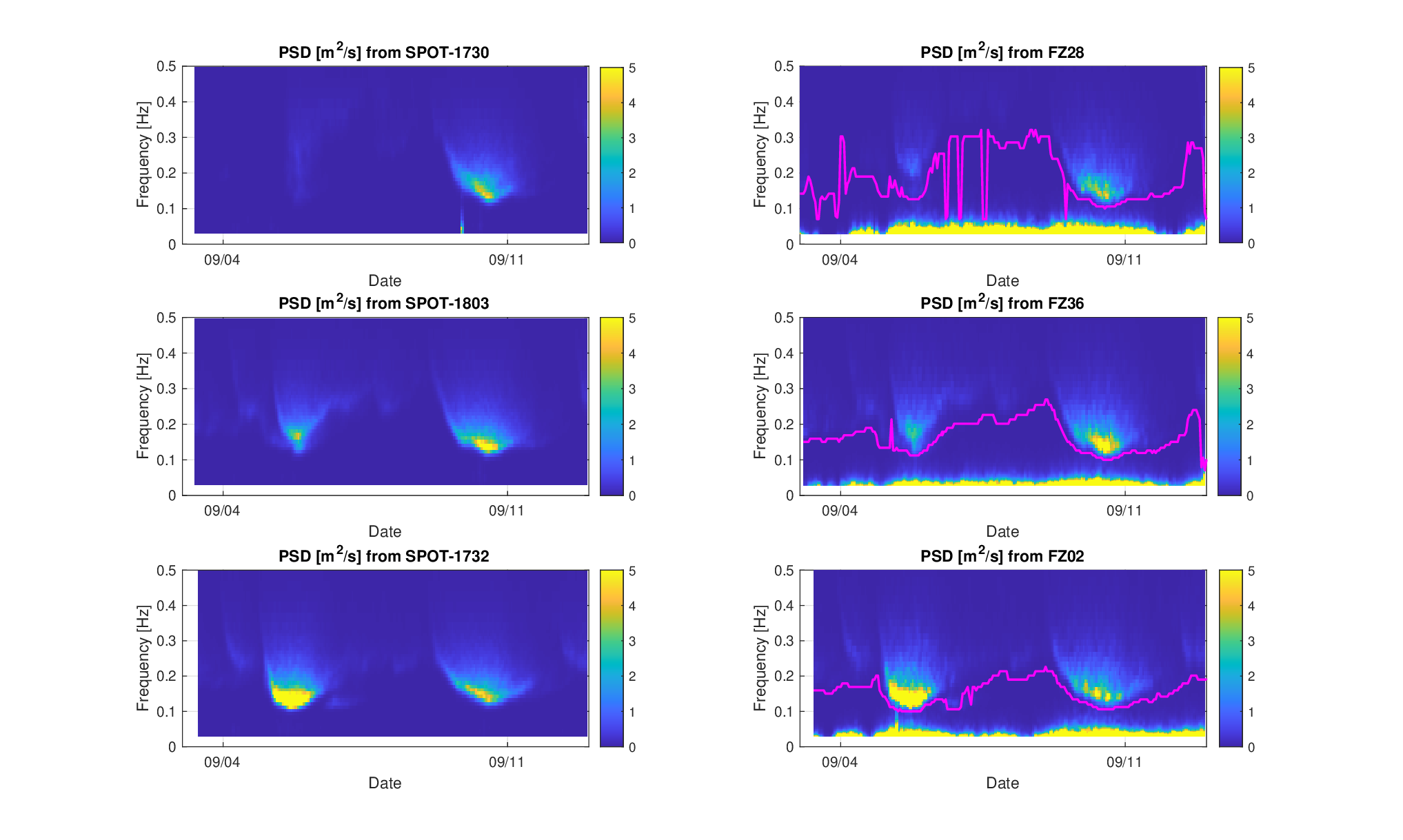}}
\caption{
Power spectral densities obtained from the three pairs of Spotter and XFZ wave buoys. The magenta line in the panels for XFZ indicates the cut-off frequency for the calculation of the bulk wave statistics.} 
\label{figure11}
\end{figure}

The frequency wave spectra obtained from XFZ and Spotter wave buoys are shown in Figure~\ref{figure11}. Despite the fact that both wave buoys transmit data hourly, the duration of the wave measurement is different. XFZ utilizes a time series of 1024 seconds, while Spotter utilizes a time series of 3600 seconds to compute power spectral density (PSD). To account for this difference in measurement duration, the frequency spectra are averaged over 2 hours for comparison. The results show that XFZ effectively measured the frequency power spectra. As mentioned in a previous study \citep{Rabault2022}, a significant difference in the lower frequency range is likely due to the measurement noise of the accelerometer. For this reason, as described in Method section, an ideal high-pass filter was applied by specifying the frequency where the power spectral density takes the local minimum (see, the magenta lines in Figure~\ref{figure11}).

\begin{figure}
\centering
\resizebox*{14cm}{!}{\includegraphics{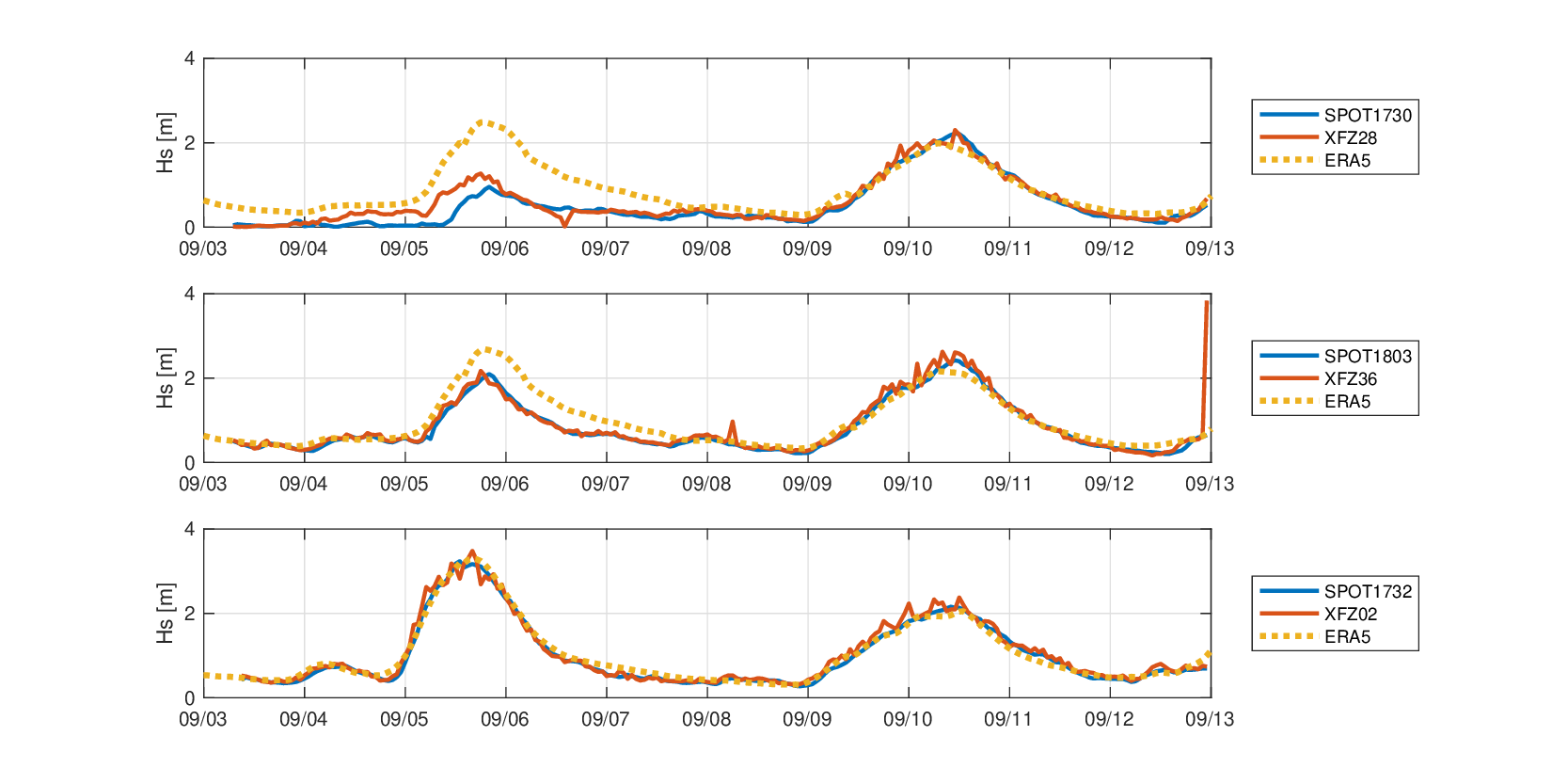}}
\caption{
Time series of the significant wave height for the three pairs of XFZ and Spotter wave buoys.} 
\label{figure8}
\end{figure}

\begin{figure}
\centering
\resizebox*{14cm}{!}{\includegraphics{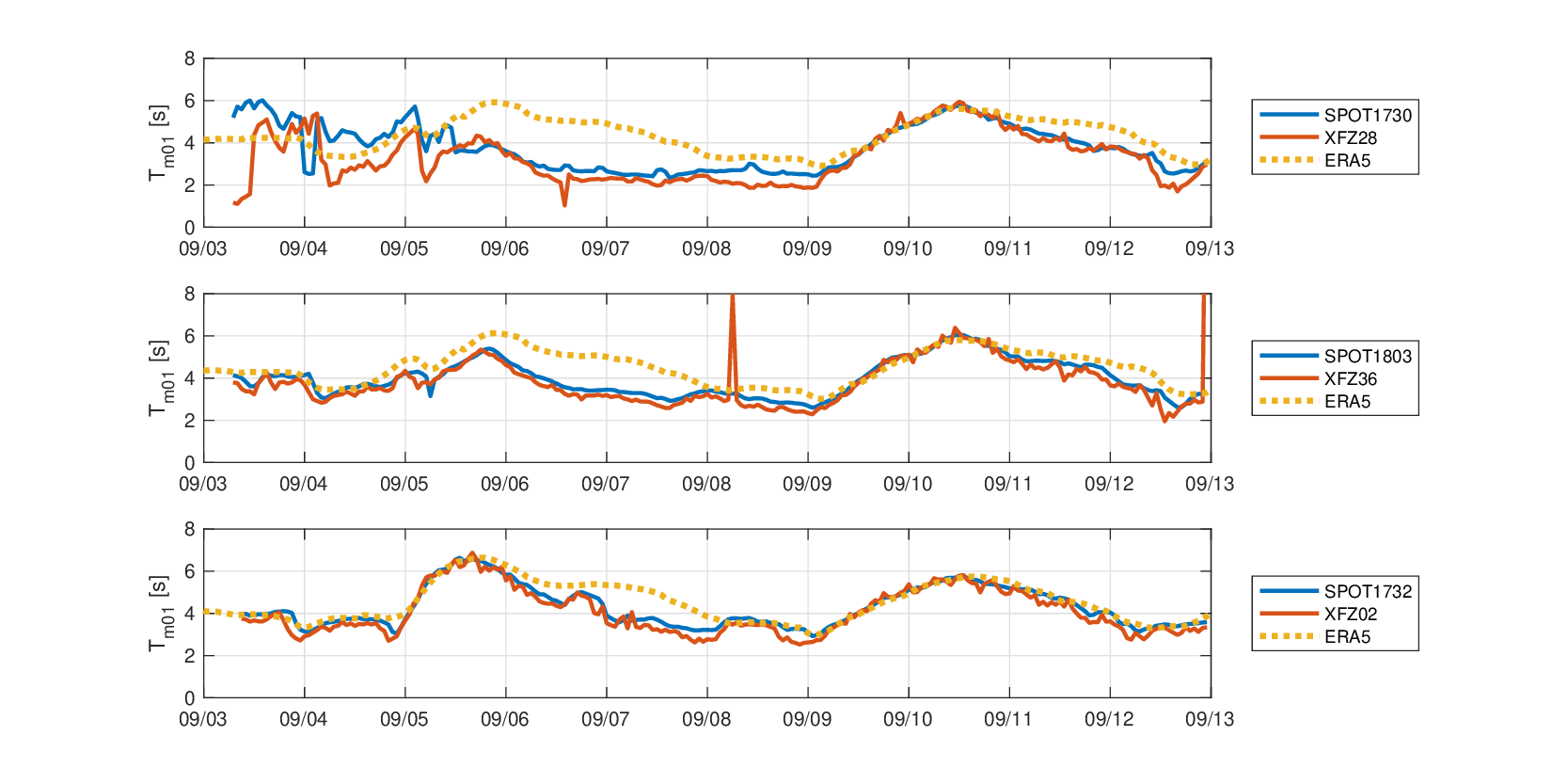}}
\caption{
Time series of the mean wave period for the three pairs of XFZ and Spotter wave buoys.} 
\label{figure9}
\end{figure}

\begin{figure}
\centering
\resizebox*{12cm}{!}{\includegraphics{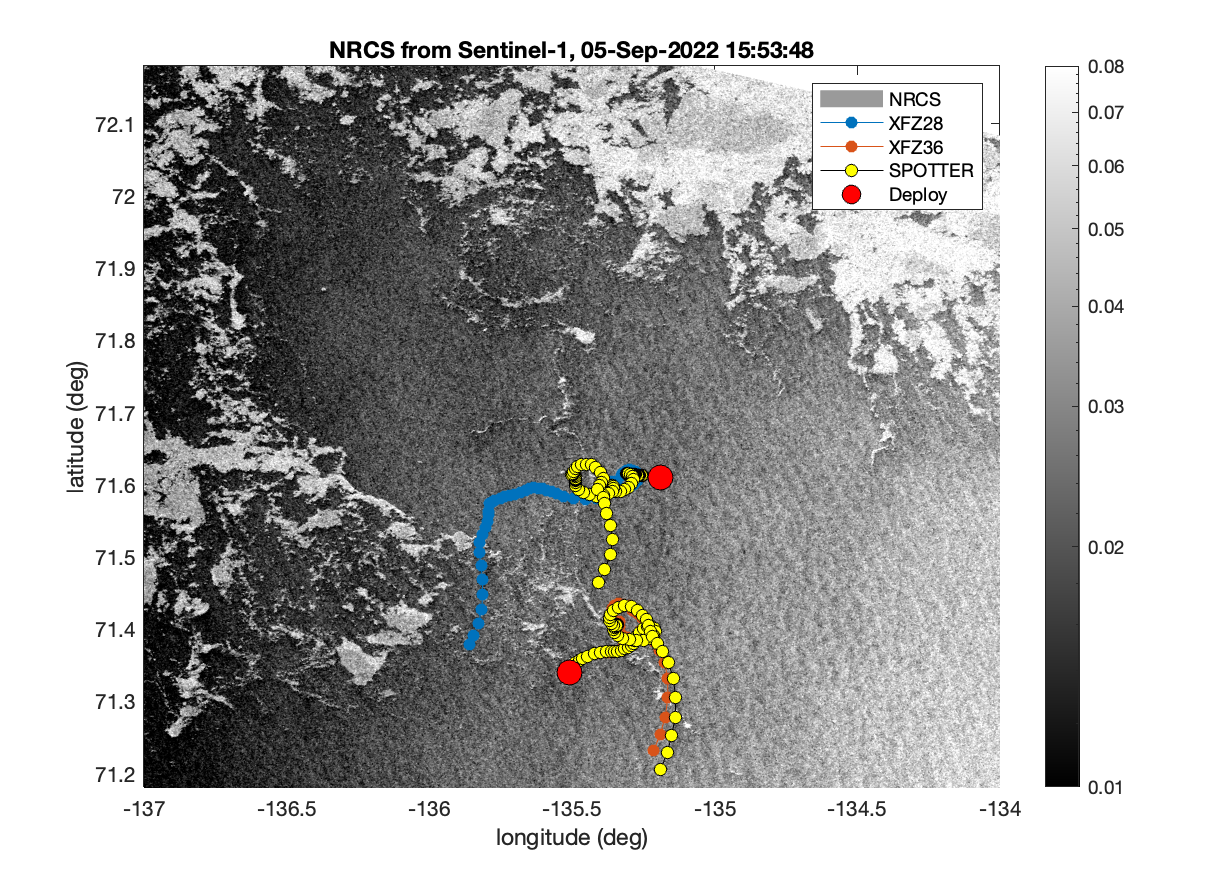}}
\caption{SAR image with the trajectories of the two pairs of XFZ and Spotter wave buoys. The trajectories are shown by the time when the SAR image was captured by Sentinel-1 on September 5th, 2022} 
\label{figure10}
\end{figure}

Figures \ref{figure8} and \ref{figure9} display the time series of significant wave heights and mean wave periods obtained from the three pairs of wave buoys and estimates from ERA5 reanalysis. Between the XFZ and Spotter wave buoys, a good level of agreement is observed for both parameters. A relatively large discrepancy is observed for the buoy pair consisting of SPOT1730 and XFZ28 in the early stages of their deployment. This difference is likely attributed to the spatial variability of the wave field, as the separation distance between these buoys increased rapidly within two days after deployment. The SAR image captured on September 5th and the buoy trajectories suggest that the trajectory of SPOT1730 diverged from that of XFZ28, resulting in an increase in their separation distance to nearly 20 km within a day (Figure~\ref{figure10}).

 A noticeable difference between the wave buoy measurements and ERA5 estimates are found from the September 5th for both parameters of significant wave heights and mean wave periods. A relatively high wind condition occured on the September 5th with the maximum wind speed of 13 m/s. The wind started blowing from the northwest and gradually veered to the north on the 6th and the speed decreased to 5 m/s. ERA5 simulates waves when the sea ice concentration is less than 30\% \citep{Prat2020}, which is likely the reason to fail in reproducing the waves under the off-ice conditions in the vicinity of the sea ice edge. In addition, based on the SAR image (Figure~\ref{figure10}), the sea ice edge around the wave buoys is far from the straight line and shows a complicated distribution of the sea ice covered area. The higher spatial resolution and the wave ice interaction parameterization should be necessary for the wave model to reproduce the waves observed by the wave buoys.

\begin{figure}
\centering
\resizebox*{14.cm}{!}{\includegraphics{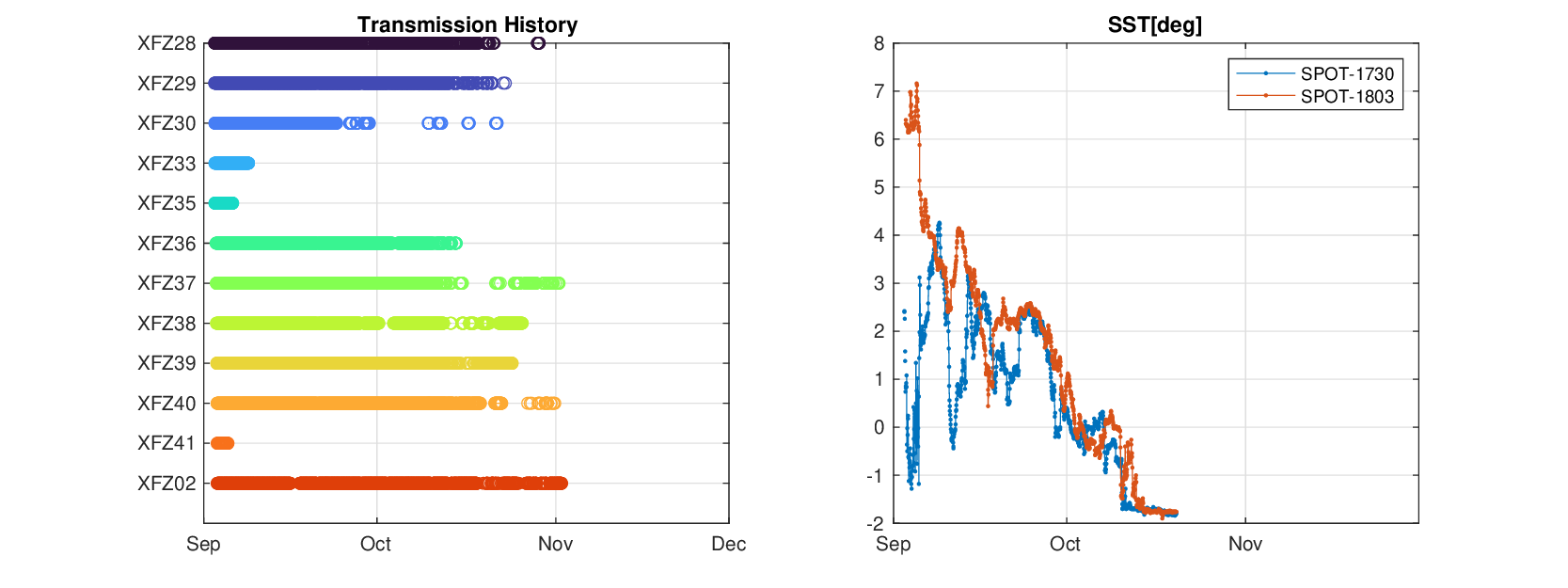}}
\caption{
 Transmission history of the twelve XFZ wave buoys deployed in the Beaufort Sea (left). Time series of the sea surface temperature measured by SPOT-1730 and SPOT-1803 (right).
} 
\label{figure12}
\end{figure}

Following the deployment of the twelve XFZs on September 3rd, three wave buoys ceased transmitting data within a week for reasons that remain unknown (Figure~\ref{figure12}). The remaining wave buoys transmitted data until mid-October but experienced intermittent transmission thereafter, with most of them eventually ceasing transmission by early November. Based on the SST measurements from the Spotter wave buoy, the intermittent transmission began when the SST decreased to the freezing temperature. The relatively small physical size of the XFZs likely led to icing on the hull and sensor box, which interfered with data transmission. Additionally, voltage drops may have occurred during periods of large current draw from the batteries, which were required to power the Iridium module. These drops may have been significant enough to cause the microcontroller to stop functioning.

\subsection{Deployment of the sensor box on ice floe}
\begin{figure}
\centering
\resizebox*{10cm}{!}{\includegraphics{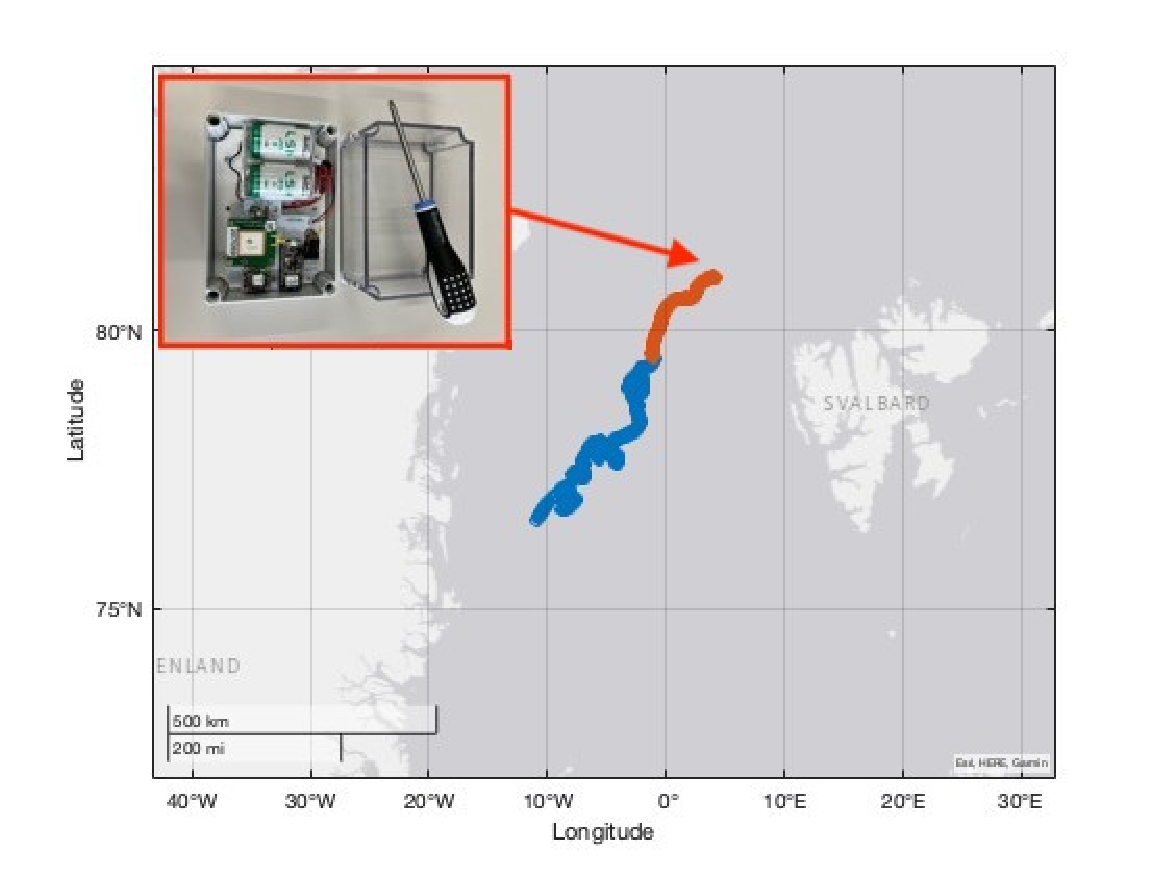}}
\caption{
The trajectories of the XFZ12, deployed on an ice floe near Svalbard, are shown. The blue line indicates the complete trajectory, but it overlaps with partial trajectories. The red line represents the trajectory from August 1st to August 15th. In the inset panel, a picture of the XFZ variant is displayed.} 
\label{figure2}
\end{figure}

As part of the Polarstern expedition (PS131) that took place between July and August 2022, a modified version of the XFZ was utilized to conduct measurements on an ice floe in the marginal ice zone (MIZ) of the Greenland Sea, situated to the northwest of Svalbard. As the device was to be placed on an ice floe, a larger watertight box was used instead of the buoy hull (see the inset in Figure~\ref{figure2}). 

This specific variant of the device, referred to as XFZ12, was equipped with a data logging function using an SD-card module compatible with the microcontroller, as it was planned to be recovered during the expedition. This deployment provided an excellent opportunity to evaluate the device's performance in measuring waves under sea ice, which typically exhibit a peak wave period exceeding ten seconds and a significant wave height less than one meter.

\begin{figure}
\centering
\resizebox*{12cm}{!}{\includegraphics{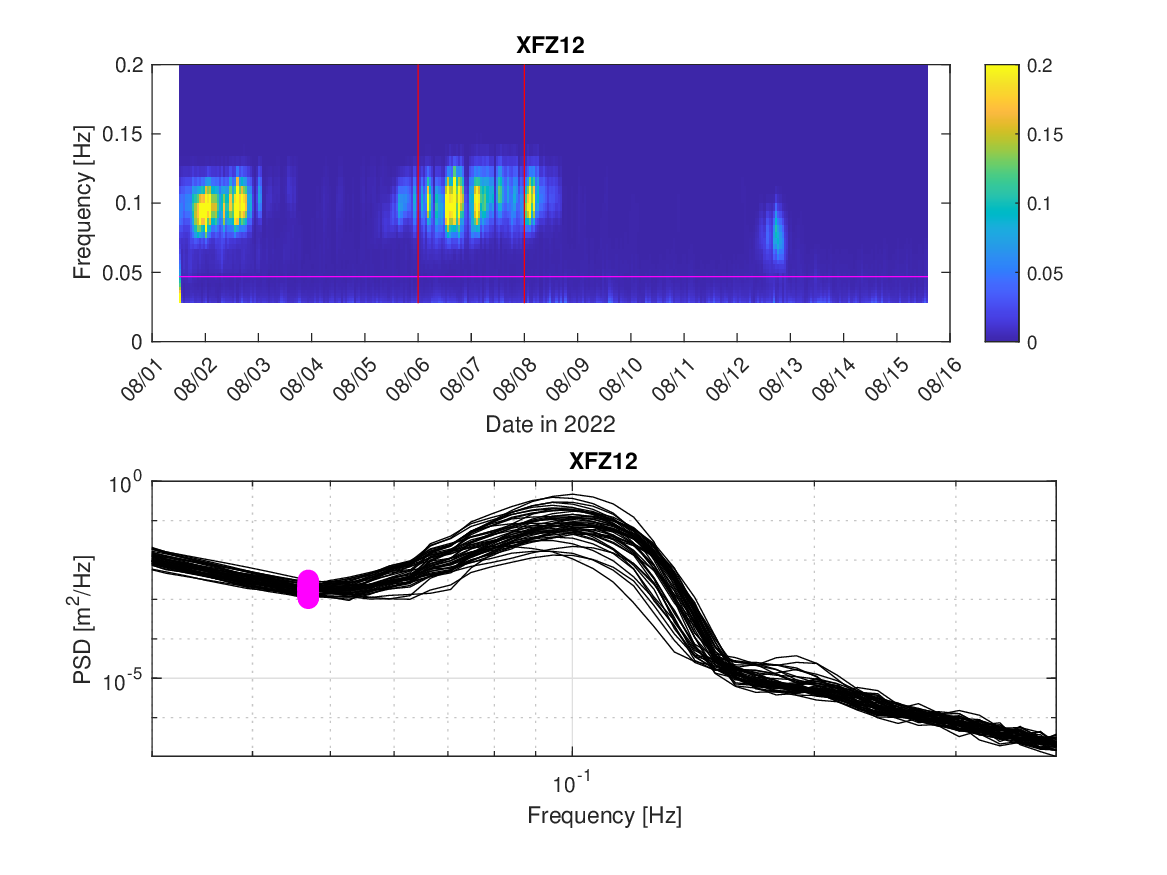}}
\caption{
(top) Power spectral densities for the period between August 1st and August 15th. The magenta line indicates the manually set cut-off frequency. (bottom) Power spectral density for the duration marked by the two red lines in the top panel.} 
\label{figure3}
\end{figure}

On August 1st at 80°45.787’N, 4°19.151’E, we deployed the sensor box on an ice floe as large as 50 m by 50 m, with an estimated thickness of approximately 2 m. Multiple wave events with the significant wave height up to 0.4~m were detected in two weeks (Figure~\ref{figure3}). The dominant frequency was observed to be around 0.1Hz. The estimated power spectral density consistently showed a single peak close to 0.1 Hz. Frequencies beyond 0.2 Hz exhibited barely any energy, suggesting that the surface waves undergo considerable scattering or attenuation if their frequency is greater than 0.2 Hz. The reason for such scattering may be the size of the ice floe, which can significantly scatter surface waves shorter than 50m if the floe size is 50m x 50m. In fact, based on the linear dispersion relation for deep water waves, the frequency of a wave 50m long would be 0.18Hz according to the deep water approximation.

The final transmission from XFZ12 occurred on 08-Oct-2022 at 01:00:00, 68 days after its second deployment. Assuming the two 13Ah batteries were fully depleted, the average power consumption per measurement is estimated to be 16.2mAh.

\section{Discussion}
The developed expandable wave buoy was deployed in the Pacific to conduct wave measurements during Typhoon Meari. The measured significant wave height did not align well with the ERA5 reanalysis data. To further investigate why ERA5 failed to reproduce the waves accurately, we conducted a comparison at the buoy's location between the JMA-MSM wind with a significantly higher grid resolution (~5km) than ERA5 (~30km). As a result, the underestimation of the significant wave height by ERA5 appears to be plausible due to a concurrent underestimation of the wind by approximately 10~\%. The timing of the peak wind speed was however less different between JMA-MSM and ERA5.

Notably, Typhoon Meari exhibits a unique characteristic of weak intensity, evident from its relatively high minimum sea surface pressure of 996 hPa. This weak intensity could have contributed to the difficulty in accurately reproducing the typhoon center and associated winds. These results indicate that direct wave measurements obtained during typhoon conditions provide valuable insights for improving numerical wave forecasting under the typhoon conditions.

XFZ was also deployed at twelve different locations in the open waters of the Beaufort Sea in September 2022. In three of these deployments, the Spotter wave buoy was deployed alongside XFZ. The frequency wave spectra obtained from XFZ demonstrated in  general a favorable agreement with the Spotter wave buoy measurements. However, some differences in the results were observed, likely attributed to the variation in measurement duration. Specifically, XFZ recorded data for slightly less than 20 minutes every hour, while Spotter maintained measurements for an hour.

Despite the discrepancy in measurement duration and consequently the results, the setup employed still yielded valuable observational data regarding the waves in the vicinity of the sea ice-covered area. This is particularly crucial, considering that existing numerical wave reanalysis products exhibit notable disparities and difficulties in accurately reproducing waves affected by the presence of sea ice, as evident from Figures \ref{figure8} and \ref{figure9}.

Results of field trials indicated that the XFZ wave measurement buoy met the fundamental requirements of buoyancy, water resistance, and durability. Nonetheless, enhancements are necessary for measurements in the marginal ice zone, as data transfer from XFZs positioned in the Beaufort Sea ceased in the period of sea ice formation. Considering that the similarly shaped but slightly larger-sized Spotter wave buoy successfully maintained data transmission during the sea ice freezing up period \citep{Kodaira2021}, a straightforward improvement would involve enlarging the buoy hull and repositioning the antenna for an Iridium SBD to a higher location. These improvements have been implemented in XFZ-V2 although the size of the wave bouy was only slightly increased from 20~cm to 24~cm (Figure~\ref{figure0}).

Regarding the customizability of the XFZ, made possible by the use of rapid prototyping technologies, was functional in improving both the hardware and firmware designs. The way to combine 3D printed objects with watertight boxes eliminated the need to test water tightness every time the hardware was redesigned. Additionally, the firmware was rapidly developed, tested, and improved using embedded programming language, further enhancing the customizability. Overall, our concept of the affordable and customizable wave buoy was proofed. From the long term point of view, the concept of an affordable and customizable device is also essential for the development in future ocean monitoring systems where sensors are utilized extensively and diversely to meet global and local demands.

During our implementation of the XFZ wave buoy using an affordable 3D printer, we have discovered certain limitations. One such limitation pertains to scalability. While it is possible to increase the number of buoys by using multiple affordable 3D printers, the production speed was limited by the 24-hour printing time, thereby making it a less scalable solution. Furthermore, the maintenance of the printer is often required, and the printing success rate is not always high. A better solution would be to build a mold for a mass production. Another limitation of the affordable 3D printer is the size of the object that can be printed. While the latest affordable printer has a twice larger maximum print size, the size of the printable object remains a limitation for the implementation method used in this study. This limitation becomes more significant when adding more sensors and batteries to the buoy. It appears rational and efficient to shift at some point towards mass production, utilizing techniques such as injection molding, subsequent to the trial-and-error phase aimed at refining and concluding the wave buoy design.

\section{Summary and Conclusion}
A cost-effective and customizable wave buoy, the XFZ, has been developed for deployment in the marginal ice zone. The design of the device is based on the use of rapid prototyping technologies, specifically the FDM-type 3D printer and the programmable microcontroller, to create a functional device that is low-cost and easily customizable. Compared to other wave buoys, the XFZ is incredibly lightweight, weighing only 2.2kg and having a volume of 4.5L. The effectiveness of the design concept was confirmed through actual manufacturing and field trials, which included deployments on an ice floe near Svalbard, measurement of typhoon conditions, and deployment of twelve XFZs in open waters in the Beaufort Sea.

The accuracy of the wave measurements obtained from the developed XFZ buoy was assessed by comparing it with the measurements from the Spotter wave buoys deployed alongside the XFZs in the Beaufort Sea. The results showed a good agreement in the significant wave heights, mean wave periods, and power spectra. Although there were some instances of missing data, two XFZs deployed in typhoon conditions managed to survive. These findings suggest that the XFZ is an affordable and durable wave buoy. To make the buoy more effective in the marginal ice zones, some improvements are necessary to continue measurements over the period of sea ice formation.

The limitations of the current implementation approach have been outlined and attributed to the utilization of rapid prototyping technologies of 3D printers for manufacturing. In essence, scalability and the size of the buoy are constrained by the presently available affordable printers. As a result of these limitations, our approach is presumed to be advantageous for small research groups seeking to explore a novel ocean sensing technique, but may not be a suitable option for those planning large-scale sensor deployments.

In terms of future work, it is important to consider the measurement of directional wave spectra. However, due to the weak horizontal magnetic field in polar regions, the heading information obtained from the 9-axis IMU may not be reliable. An alternative approach is to use GPS fixes to obtain horizontal displacement and combine it with vertical displacement obtained from IMU measurements. Additional sensors such as an ultrasonic anemometer could also be added to expand the range of measurement items, as wind information is crucial in understanding wind wave development. Moreover, recent research \citep{Rogers2016} has suggested that visual information from a digital camera can be useful for interpreting observational results in the marginal ice zone, thus highlighting the potential value of this technology for future studies.

\section*{Acknowledgement}
We are grateful to the crews, and scientists on board R/V Mirai during the MR22 Arctic expedition for supporting our buoy deployment. We are also grateful to the crews, and scientists on the Polarstern during the expedition (PS131) for supporting the deployment of wave sensing device. ERA5 reanalysis was obtained using Copernicus Climate Change Service Information. The Level-2 SAR data of the normalized radar cross-section from Sentinel-1B was used to estimate the sea ice extent near the study area. The data were created by NOAA and were obtained from NOAA CoastWatch (http://coastwatch.noaa.gov). The original data were provided to NOAA by the Copernicus Program.

\section*{Funding}
This work was supported by the Japanese Ministry of Education, Culture, Sports, Science, and Technology through the Arctic Challenge for Sustainability II (ArCS II) Project (Program Grant Number JPMXD1420318865); JSPS under KAKENHI grant  19H00801, 19H05512, 21K14357, and 22H00241.

\section*{Disclosure statement}
Any financial conflict with the subject matter discussed in the manuscript is completely disclosed in the acknowledgment section of the manuscript.

\section*{Supplemental online material}
The CAD file of the hull design described in this study is provided as supplemental online material. 

\bibliographystyle{tfcad}
\bibliography{export}

\newpage


\newpage

\appendix
\section{Wave buoy production}
\subsection{Hardware (hull)}
The fundamental design of the hull follows the wave buoy that was introduced in our previous study \citep{Kodaira2022}. In order to realize our concept, we used Rapid Prototyping technology with a Fused Deposition Modeling (FDM) type 3D printer. One issue with 3D printed products for use in ocean-sensing devices is the lack of watertightness. To address this problem and protect the electronics, we employed commercial watertight boxes with International Protection (IP) rating IP66/67 (SPCP081308T and SPCP101004T, TAKACHI ELECTRONICS ENCLOSURE CO., LTD) in combination with the 3D printer. The hull is designed to take the form of a shell and enclose the boxes within (see Figure A1 a). This combination of a watertight box and 3D printer enables us to create something that is watertight while still maintaining high customizability.

The power source for our prior wave buoy FZ was solar energy. However, for the polar oceans where solar radiation is limited, Lithium metal primary batteries were selected as the optimal energy source. The batteries were safely contained within a discrete watertight enclosure and supplied power to the electronics (illustrated in Figure A1b). The two-box architecture was specifically designed to avoid the top heavy structure by positioning the battery box at a lower level while maintaining the sensor box at a higher position. The cable connection between the boxes was made watertight using a cable gland with an IP rating of IP67. The new wave buoy has been designated as eXpendable FZ (XFZ) due to the implementation of primary batteries.

The 3D printer was utilized to effectively integrate the above-mentioned components. Guides and mounting points were fabricated on the hull to facilitate the assembly of the watertight boxes. Holes were prepared to attach an eyebolt that serves as a contact point for the external weight (as shown in Figure A1a). Attaching the external weight is a simple yet effective solution to lower the gravity center below the metacenter to ensure floating stability and prevent possible turnover during the measurements.

A suitable buoyant material is necessary to fill the void between the hull and watertight boxes. Considering its durability and extensive use in marine applications, polyurethane foam (PUF) was deemed appropriate. Two-part liquid foam (469-95IK4, PROST) was utilized, which expands up to 10 times its original volume and sets rigid once mixed in equal parts. The expanded foam has a nominal density of approximately 100 kg/m³. By filling the void with PUF, the wave buoy can achieve rigid buoyancy.

The assembly procedure was as follows. The battery box was initially positioned inside and the two liquid components of Polyurethane Foam (PUF) were manually mixed by hand and poured in, as illustrated in Figure A1c. The sensor box was subsequently placed on top of the buoy hull, and the two liquids inside the hull initiated a gradual expansion, eventually filling the vacant space within the buoy, as depicted in Figure A1d. Lastly, a 0.5 meter-long stainless steel ballast chain was affixed to the buoy, as shown in Figure A1e.

The assembled buoy has a weight of 2.2 kilograms and a volume of 4.5 liters, with the waterline resting in the center of the hull, as illustrated in Figure A1f. Its diameter measures approximately 20 centimeters. In contrast to the prior wave buoy FZ, the XFZ is heavier, as it includes primary batteries. To enhance buoyancy, the volume inside the hull was increased by slightly expanding the dimensions and designing the buoy in a more rounded shape. The detailed design can be accessed from the provided CAD file, which is included as supplementary information.

A stress test was conducted for the produced buoy hull using the compression testing machine (AG-XD plus, 20 kN - 50 kN, SHIMADZU CORPORATION). The force was applied on the side walls of the hull, and gradually increased with the compression speed of 10 mm/s. The buoy hull is finally cracked and deformed when $1.2\times10^4$ N force is applied.

\begin{figure}
\centering
\resizebox*{11cm}{!}{\includegraphics{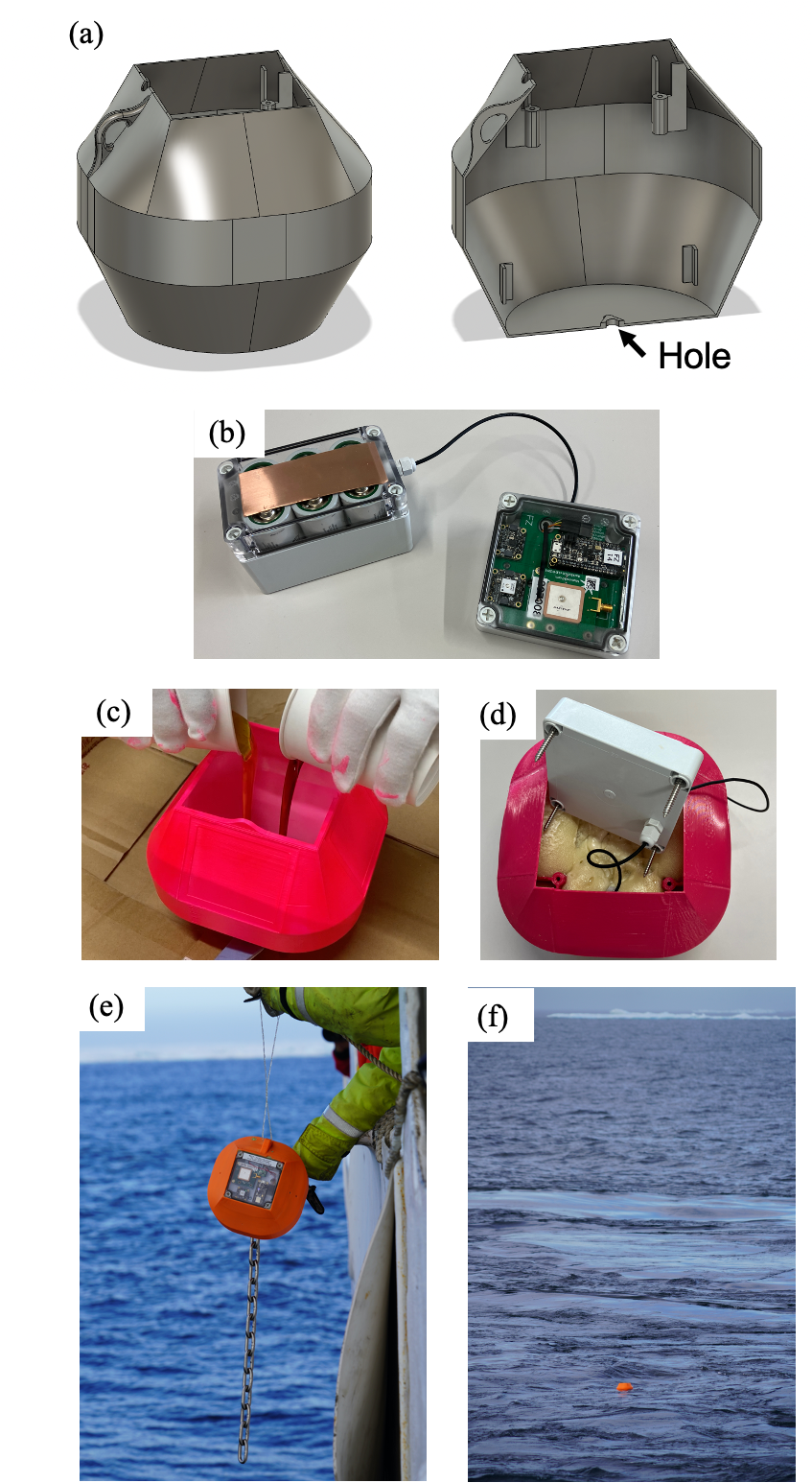}}
\caption{
(a) Computer-aided design of the hull for the wave buoy, (b) The sensor and battery boxes prior to assembly onto the eXpendable FZ (XFZ) wave buoy, (c) Introduction of two liquid components of polyurethane foam (PUF) into the hull, (d) PUF foam filling the internal volume of the hull after expansion, (e) Deployment of the wave buoy from R/V Mirai in September 2022, and (g) The wave buoy soon after deployment in the Beaufort Sea in September 2022.
} 
\label{figure1}
\end{figure}

\subsection{Electronics}

The Adafruit Feather nRF52840 Sense was carefully chosen as the microcontroller to oversee the measurement system. Some of the sensors on the board include the SHT Humidity and BMP280 temperature and barometric pressure sensors, which are utilized to monitor the environmental conditions within the sensor box. In addition to these sensors, the measurement system also comprises primary electronic devices such as the global navigation satellite systems (GNSS) receiver, Inertial Measurement Unit (IMU), and Iridium Short Burst Data (SBD) modules, all of which are housed within a watertight enclosure (please refer to Table A1 for a comprehensive list of electronics used). The power source for this measurement system is supplied by six Lithium D-cells (SAFT LSH20 or equivalent) batteries, each with a nominal capacity of approximately 13 Ah, which are utilized in parallel.

To minimize power consumption, the sensors were set to sleep mode when not in use. The corresponding energy consumption for each process was measured, and the results are presented in Table~\ref{table1}. The total energy consumption for one measurement cycle was estimated to be 17mAh. A comparison of the results with the study conducted by \citep{Rabault2022} suggests that a possible way to achieve lower energy consumption is to replace the current IMU sensor. We employed the BNO055 for IMU measurements, which appears to consume more energy compared to other IMU sensors such as the ISM330DHC. This could be due to the BNO055's onboard Euler angle calculation.

The endurance of the device in low temperatures was tested by subjecting the sensor box to a temperature of -18°C in a freezer. Despite the inability to perform GPS measurements and Iridium satellite communication under these conditions, the device exhibited expected performance for slightly over three weeks, powered by a single Lithium D-cell battery.

To investigate the ability of the IMU sensor to detect small-amplitude wave signals, as per \cite{Rabault2022}, the sensor box was placed on the wave maker in the wave-ice tank at the Kashiwa campus of the University of Tokyo in Japan. Due to the lack of satellite communication capability, data logging was performed using an SD card. A periodic wave with a 1 cm amplitude and a period of four seconds was detected with high clarity in the frequency wave spectra, exhibiting a signal-to-noise ratio of 100.

\begin{table}
\tbl{Specification of the electronics for XFZ and energy consumption of each measurement step.}
{\begin{tabular}{lccc} \toprule
 Sensor & Product & Energy consumption per cycle
 \\ \midrule
Microcontroller	& Adafruit Feather nRF52840 Sense	& 3 mA \\
IMU	& Adafruit BNO055 Absolute Orientation & 7 mAh / 1024 sec \\
GNSS receiver	&Adafruit Mini GPS PA1010D	& 1 mAh/ 120 sec \\
Iridium module	&Rockblock 9603N	& 4 mAh/ 120 sec \\
SDcard module (optional) & Adalogger FeaterWing - RTC + SD Add-on	&5 mAh / 1024 sec \\
DC-DC converter	&Adafruit MiniBoost 5V @ 1A -TPS61023	& - \\ \bottomrule

\end{tabular}}
\label{table1}
\end{table}

\end{document}